\RequirePackage{fix-cm}
\documentclass[singlecolumn,epjc3]{svjour3}
\smartqed  
\RequirePackage{graphicx}
\usepackage{graphicx}
\usepackage{graphicx,amsfonts,amssymb}
\usepackage{amsmath,amssymb,enumerate}
\usepackage{placeins}
\usepackage{multirow}
\usepackage{multicol}
\usepackage{float}
\journalname{Eur. Phys. J. A}

\begin{document}
\title{A new exact anisotropic solution of embedding class one}
	
	\author{S.K. Maurya\thanksref{e1,addr1}
		\and Y.K.Gupta\thanksref{e2,addr2}
		\and Smitha T. T. \thanksref{e3,addr1}
		\and Farook Rahaman \thanksref{e4,addr3}.}
	
	\thankstext{e1}{e-mail: sunil@unizwa.edu.om}
	\thankstext{e2}{e-mail:kumar001947@gmail.com}
	\thankstext{e3}{e-mail: smitha@unizwa.edu.om}
	\thankstext{e4}{e-mail: rahaman@iucaa.ernet.in}

	\institute{Department of Mathematical \& Physical Sciences, College of Arts \& Science,
		University of Nizwa, Nizwa, Sultanate of Oman\label{addr1}
		\and Department of Mathematics, Raj Kumar Goel Institute of Technology, Ghaziabad, 201003, U.P. (INDIA) \label{addr2}
		\and Department of Mathematics, Jadavpur University, Kolkata 700032, West Bengal, India\label{addr3}}
		
	\date{Received: date / Accepted: date}
	
	\maketitle	
\begin{abstract}
\noindent We have presented a new anisotropic solution of Einstein's field equations for compact star models. The Einstein's field equations are solved by using the class one condition \cite{1}. We constructed the expression for anisotropy factor ($\Delta$) by using the pressure anisotropy condition and there after we obtained physical parameters like energy density, radial and transverse pressure. These models parameters are well behaved inside the star and satisfy all the required physical conditions. Also we observed a very interesting result that all physical parameters are depending upon the anisotropy factor ($\Delta$). The mass and radius (Tab. 1) of the present compact star models are quite compatible with the observational astrophysical compact stellar objects like Her X-1, RXJ 1856-37, SAX J1808.4-3658(SS1), SAX J1808.4-3658(SS2).
	\end{abstract}

\keywords{General relativity, metric functions, anisotropic factor and Compact stars.}

\section*{I. INTRODUCTION:}

\noindent It is still an open problem in astrophysics to understand the nature and exact composition of compact stars which are more compact than ordinary neutron stars and therefore, it has become a field of active research in recent years.  Now, it is an obvious question for astrophysicists that what would be the pertinent back ground space-time to model such class of compact stars. Recently, Avellar and Horvath \cite{2} have obtained a various set of exact and approximate solutions to model strange stars. The plan of this investigation is to look for a substantially feasible logical model which can designate such class of compact stars. Bowers and Liang \cite{3} highlighted on the significance of locally anisotropic equations of state for relativistic fluid sphere models. Herrera and Santos \cite{4} have discussed effect of local anisotropy in pressure and proposed several physical mechanisms in low and very high density system for astrophysical compact objects. Also recently L. Herrera et al.\cite{42} have discussed a general study on the spherically symmetric anisotropic fluid distribution.

Several investigations \cite{5,6,7,8,9,10} showed that anisotropy effects should be considered on such parameters like maximum equilibrium mass and surface red-shift. Ruderman \cite{11} has studied  the stellar  models and proposed  that the nuclear matter (where the nuclear interaction  must be treated relativistically)  may have anisotropic structures   for  high density ranges  of order $10^{15}gm/cm^{3}$. Recent observations on highly compact stars indicate that the densities of such objects exceed the nuclear matter density. Also, theoretical advances in recent time argue that pressures within such stars are anisotropic in such bodies Sharma and Maharaj \cite{12}. Abreu et al. \cite{13} have discussed the influence of density fluctuations and local anisotropy has on the stability of local and non-local anisotropic matter configurations by using concept of cracking. Thus anisotropy have been introduced as an important characteristic in the formation of the compact stars which allowed several authors to model the objects substantially more practical \cite{14,15,16}. Recently Maurya et al. \cite{17} has also proposed an algorithm for all spherically symmetric anisotropic charged compact star.  Further investigations of compact stars some other distinguished works with different aspects are as follows \cite{18,19,20,21,22,23,24,25,26,43,44,45,46}.\\
 \noindent In the present article, we have obtained anisotropic compact star models in class one metric. The beauty of class one condition is that the metric functions $\lambda$ and $\nu$ are dependent to each other. Due to such relation, if anisotropy is zero then we will get only two types of perfect fluid solutions either Schwarzschild solution \cite{27} or Kholar Chao solution \cite{28}. For this purpose we have started with the metric functions which is not same as Schwarzschild or Kohlar Chao metric function and obtained expressions for anisotropy factor, energy density and pressures.  In this way we got a very interesting result that the radial and transverse pressure and energy density are dependent on anisotropy factor i.e. if anisotropy vanishes identically then pressures and density vanishes automatically and metric turns out be flat.This implies that interior of star may be pure anisotropic.
\\ The contents of the article as follows: The Sec. II, we divided in two subsections as (a) and (b): In (a). we set up the Einstein's field equation for anisotropic fluid distributions and In subsection (b). we proposed class one condition and develop an important relation between metric potentials $\lambda$ and $\nu$ by using above class one condition. After that we obtained the expression for anisotropic factor $\Delta$ in terms of $\lambda$ and $\nu$ which gives a very important result. In Sec.III, we determine the physical parameters $\rho$,\, $p_{r}$, \, $p_{t}$ and $\Delta$  form obtained $\lambda$ and $\nu$.Also in this section, we determined two generating function for the anisotropic solution by using Herrera et al. \cite{54} algorithm. The Sec.IV contains the physical features of the models, the details as follows (a). regularity at centre, (b). velocity of sound conditions, (c). Behavior of equation of state parameters $\omega_{r}$ and $\omega_{t}$,(d) matching conditions and (e). Equilibrium condition by using generalized TOV equations. The stability analysis of the models has been discussed with the help of the cracking concept \cite{29,30} in Sec.V. In the Sec.VI, we have given the relation between effective mass and radius, surface red shift and numerical values of obtained anisotropic compact stars. In Sec.VII: we presented about comparison between our present anisotropic models with the observational compact objects. At last Sec.VIII, we discussed conclusion about the present anisotropic models with some numerical data of physical parameters.

\section*{II. EINSTEIN'S FIELD EQUATIONS AND CLASS ONE METRIC:}

(a) Field Equations:\\

To describe the space time of compact stellar configuration, we consider the static spherically metric as \cite{31,32}:
\begin{equation}
ds^{2} =-e^{\lambda } dr^{2} -r^{2} (d\theta ^{2} +\sin ^{2} \theta \, d\phi ^{2} )+e^{\nu } \, dt^{2}
\label{1}
\end{equation}
Now assume the energy momentum tensor for the matter distribution, the interior matter of the star may be expressed in the following standard form as:
\begin{equation}
\label{2}
T_{ij}=diag(\rho, -p_{r}, -p_{t}, -p_{t}) ,
\end{equation}
Where, $\rho$,\,$p_r$,\, $p_t$ and corresponding to energy density, radial and tangential pressure respectively of matter distribution.
The Einstein field equation can be written as:
\begin{equation}
\label{3}
R_{i,j}-\frac{1}{2}\,R\,g_{i\,j}=-8\pi\,T_{i\,j},
\end{equation}
Here $G = c = 1$ under geometrized relativistic units.
In view of metric (1), the equation (3) gives following differential equations \cite{33}:
\begin{equation}
\label{4}
p_{r} =\frac{e^{-\lambda}}{8\pi}\left[\frac{v'}{r}  -\frac{(e^{\lambda}-1)}{r^{2}} \right]  ,
\end{equation}
\begin{equation}
\label{5}
 p_{t} =\frac{e^{-\lambda}}{8\pi}\left[\frac{v''}{2} -\frac{\lambda'v'}{4} +\frac{v'^{2} }{4} +\frac{v'-\lambda'}{2r} \right] ,
\end{equation}
\begin{equation}
\label{6}
\rho =\frac{e^{-\lambda}}{8\pi}\left[\frac{\lambda '}{r} +\frac{(e^{\lambda}-1 )}{r^{2}}\right]
\end{equation}
Pressure anisotropy condition:
\begin{equation}
\label{7}
\Delta =\,p_{t} -\, p_{r} \,=\frac{e^{-\lambda }}{8\pi}\left[\frac{v''}{2} -\frac{\lambda 'v'}{4} +\frac{v'^{2} }{4} -\frac{v'+\lambda '}{2r} +\frac{e^{\lambda }-1}{r^{2} } \right]  ;
\end{equation}

\noindent (b). Class one condition:\\
 As we know that the manifold $V_n$ can always be embedded in Pseudo Euclidian space $E_m$ of $m$ dimensions with $m = n(n+1)/2$ \cite{47}. The minimum extra dimension $K$ of the Pseudo-Euclidian space to embedded $V_n$ in $Em$, is called the class of the manifold $Vn$ and it must be less than or equal to the number $(m-n)$ or same as $n(n-1)/2$. The embedding class $K$ turns out to be $6$ in the case of relativistic space time $V_4$. In particular the class of spherical symmetric space-time is $2$ while plane symmetric is of class $3$. The Schwarzschild's exterior and interior solutions are of class $2$ and class $1$ respectively. Moreover The famous Friedman-Robertson-Lemaitre \cite{48,49,50} space-time is of class $1$ and the famous Kerr metric of class $5$ \cite{51}. The postulates of general relativity do not provide any physical meaning to higher dimensional embedding space. However, it provides new characterizations of gravitational field, which hopefully can be connected to  internal symmetries of elementary particle physics \cite{52}. Embeddings are also found connected to, extrinsic gravity, strings and membranes and new brain world (Pavsic and Tapia \cite{53}).

Now the metric (1) may represent the space time of emending class one, if it satisfies the Karmarker condition \cite{34} as:
\begin{equation}
\label{8}
R_{1414}=\frac{R_{1212}R_{3434}+R_{1224}R_{1334}}{R_{2323}},
\end{equation}
with $R_{2323}=0$ \, [35].\\

After inserting the curvature components $R_{hijk}$ of the metric (1) into equation (8), we get:

\begin{equation}
\label{9}
\frac{2\nu''}{\nu'}+\nu'=\frac{\lambda'\,e^{\lambda}}{e^{\lambda}-1}
\end{equation}
The solution of above Eq. (5) gives the following relation between $\nu$ and $\lambda$ as:

\begin{equation}
\label{10}
\nu=2\,ln\left[A+B\int{\sqrt{(e^{\lambda(r)}-1)}dr}\right].
\end{equation}
 where,  A and B  are non - zero arbitrary constant of integration.
 \noindent  Using the Eq.(10) in Eq.(7) and after well setting of Eq.(7), we get:

\begin{equation}
\label{11}
\Delta=\frac{\nu'\,e^{-\lambda}}{32\,\pi}\,\left(\frac{\nu'\,e^{\nu}}{2B^{2}r}-1\right)\,\left(\frac{2}{r}-\frac{\lambda'e^{-\lambda}}{1-e^{-\lambda}}\right).
\end{equation}

We note from Eq. (11) that if anisotropy is zero then at least one factor of right side of Eq. (11) should be zero. So if first factor is zero then corresponding solution will be Kohlar-Chao\cite{28} solution while second factor will gives the Schwarzschild solution \cite{27}.

\section*{III. INTERIOR STRUCTURE OF THE COMPACT STAR:}

\subsection*{IIIa. Anisotropic solution of embedding class one:}

Our next task to find the anisotropic solution for compact star of class one. As above argument in Sec.II, we suppose the metric potential of the form:

\begin{equation}
\label{12}
 e^{\lambda}=1+\frac{(a-b)r^{2}}{1+br^{2}}.
\end{equation}
where  $a$ and $b$ are constants with $a\neq b$.\\

The above form of metric potential is 1 at centre and positive- finite everywhere inside the star. This implies that it is free from singularity and physically valid.\\

\noindent  By plugging the Eq.(12) into (10), we get:
 \begin{equation}
\label{13}
 \nu=2\,ln\left[A\,b+B\,\sqrt{(a-b)}\,\sqrt{1+br^{2}}\right]-2\,ln\,b.
\end{equation}

The expression for energy density and pressures are given as:
\begin{equation}
\label{14}
 \rho=\frac{(a-b)}{8\pi}\left[\frac{(3+ar^{2})}{(1+ar^{2})^{2}}\right].
\end{equation}

\begin{figure}[!htp]\centering
	\includegraphics[width=5.5cm]{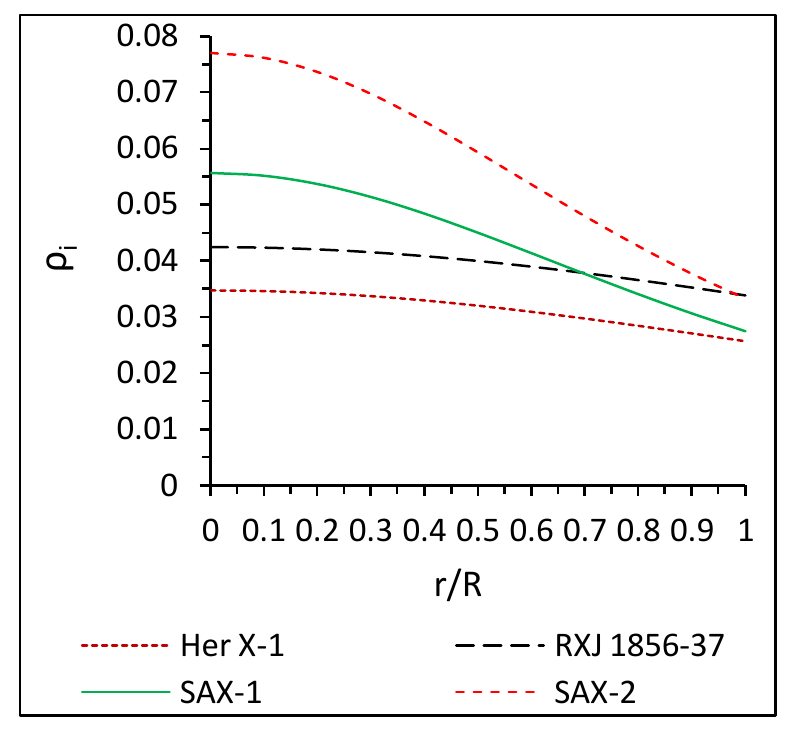}
	\caption{Variation of energy density ($\rho_i=8\pi \,\rho$) with respect to fractional radius (r/R). For this graph, the values of constants are as follows: (i). $a=0.0045, b=-0.0071, A=1.3704, B=0.0490$ with mass$(M)=0.9824 M_{\odot}$ and radius$(R)=6.7$ Km for $Her X-1$, (ii). $a=0.0041, b=-0.0101, A=1.2705, B=0.0555$ and mass $(M)=0.9042 M_{\odot}$, radius$(R)=6.0$ Km for $RXJ 1856-37$, (iii). $a=0.0011, b=-0.0075, A=1.4134, B= 0.0547$, mass $(M)=1.4349 M_{\odot}$ and radius$(R)=7.07$ Km for $SAX J1808.4-3658(SS1)$ and (iv). $a=0.0017, b=-0.0087, A=1.5404, B= 0.0617$, Mass$(M)=1.3237 M_{\odot}$ and radius$(R)=6.37$ Km for $SAX J1808.4-3658(SS2)$}
	\label{Fig1}
\end{figure}

\begin{equation}
\label{15}
 p_{r}=\frac{\sqrt{(a-b)}}{8\pi}\,\left[\frac{-A\,b\,\sqrt{(a-b)}+B\,(3b-a)\,\sqrt{1+br^{2}}}{(1+ar^{2})[A\,b+B\,\sqrt{(a-b)}\,\sqrt{1+br^{2}}]}\right].
\end{equation}\\
\begin{figure}[!htp]\centering
	\includegraphics[width=5.5cm]{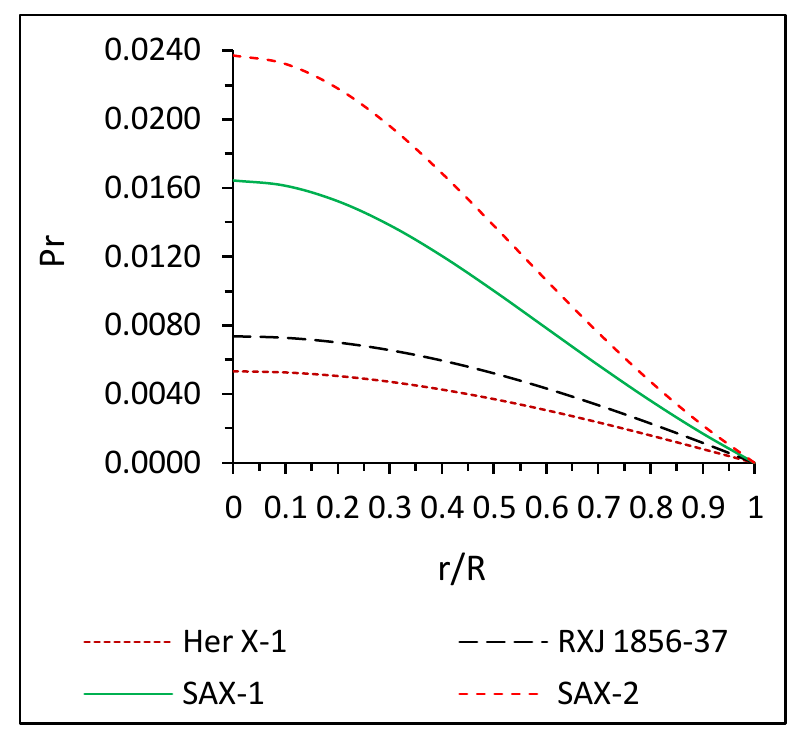}\includegraphics[width=5.5cm]{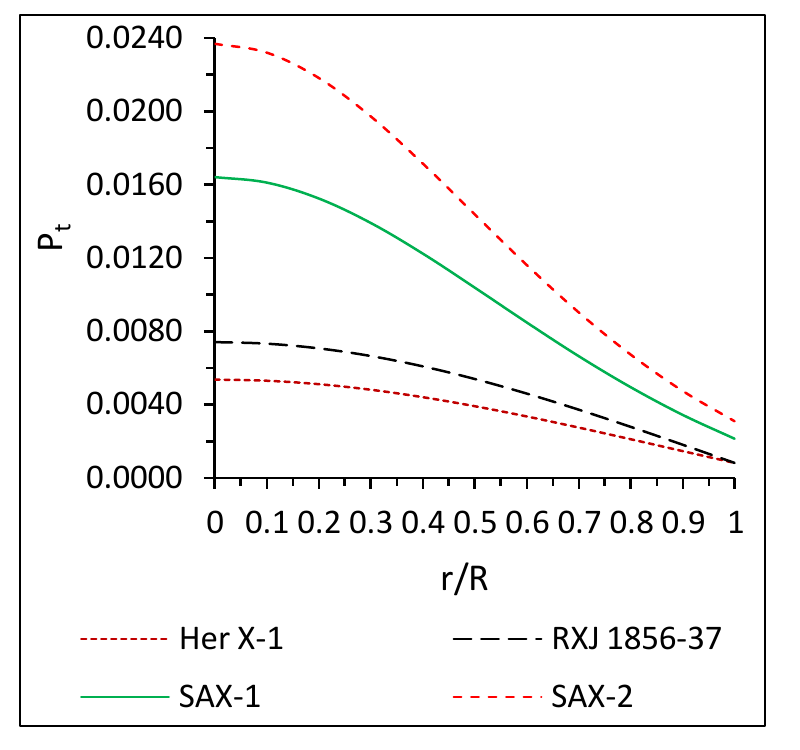}
	\caption{Variation of radial pressure (left panel) and transverse pressure (right panel) with respect to fractional
		radius (r/R). where, $P_{r}=8\pi\,p_{r}$ and $P_{t}=8\pi\,p_{t}$. For the purpose of plotting this figure, we have employed the same data set for the arbitrary constants: a, b, A and B with same mass and radius for each different compact star as in Fig. 1}
	\label{Fig2}
\end{figure}
\begin{equation}
\label{16}
 p_{t}=\frac{\sqrt{(a-b)}}{8\pi}\,\left[\frac{-A\,b\,\sqrt{(a-b)}\,\sqrt{1+br^{2}}+B\,(3b-a+ab^{2}r^{4})\,(1+br^{2})}{(1+ar^{2})^{2}\,\sqrt{1+br^{2}}\,[A\,b+B\,\sqrt{(a-b)}\,\sqrt{1+br^{2}}]}\right].
\end{equation}\\

In this model, the measure of anisotropy is determined by the Eq.(11) as:

\begin{equation}
\label{17}
 \Delta=\frac{ar^{2}\,\sqrt{(a-b)}}{8\pi}\,\left[\frac{A\,b\,\sqrt{(a-b)}\,\sqrt{1+br^{2}}+B\,(a-2b)\,(1+br^{2})}{(1+ar^{2})^{2}\,\sqrt{1+br^{2}}\,[A\,b+B\,\sqrt{(a-b)}\,\sqrt{1+br^{2}}]}\right].
\end{equation}
\begin{figure}[!htp]\centering
	\includegraphics[width=5.5cm]{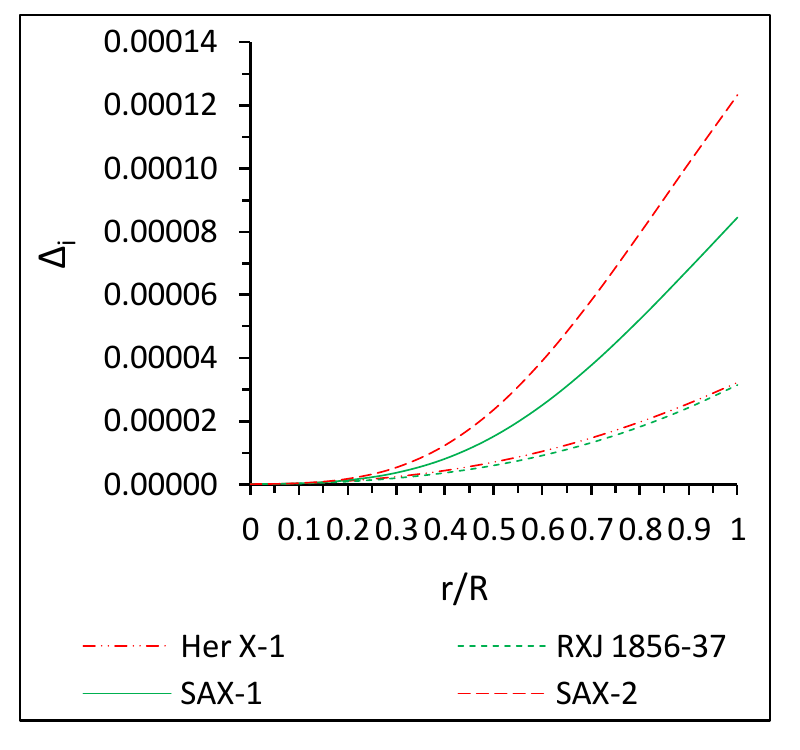}
	\caption{Variation of anisotropy factor ($\Delta_i=p_t-p_r$) with respect to fractional radius (r/R). For the purpose of plotting this figure, we have employed the same data set for the arbitrary constants: a, b, A and B with same mass and radius for each different compact star as in Figs. 1 and 2}
	\label{Fig3}
\end{figure}

We observe from Eq. (17), the anisotropy $\Delta$ is zero for every $r$ if (i) $a=0$ and (ii) $a=b$. The first case gives Schwarzschild's interior solution while second case gives flat metric and energy density, pressures will become zero . The second case gives very interesting result that total energy density, radial and tangential pressure of a compact star dependent on anisotropy factor only.\\

We propose the ratio of radial pressure and tangential pressure verses density by the function $\omega_{r}$ and $\omega_{t}$ as: $\omega_{r}=\frac{p_{r}}{\rho}$ and $\omega_{t}=\frac{p_{t}}{\rho}$.\\
Then the expressions of $\omega_{r}$ and $\omega_{t}$ for the anisotropic compact stars are given as:
\begin{equation}
\label{18}
 \omega_{r}=\frac{(1+ar^{2})}{(3+ar^{2})\,\sqrt{(a-b)}}\,\left[\frac{-A\,b\,\sqrt{(a-b)}\,+B\,(3b-a)\,\sqrt{(1+br^{2})}}{[\,A\,b+B\,\sqrt{(a-b)}\,\sqrt{1+br^{2}}\,]}\right]
\end{equation}
\begin{equation}
\label{19}
 \omega_{t}=\frac{1}{\sqrt{(a-b)}}\,\left[\frac{-A\,b\,\sqrt{(a-b)}\,+B\,(3b-a+abr^{2})\,\sqrt{(1+br^{2})}}{(3+ar^{2})\,[\,A\,b+B\,\sqrt{(a-b)}\,\sqrt{1+br^{2}}\,]}\right]
\end{equation}

\subsection*{IIIb. Generating functions for anisotropic solution of embedding class one:} Herrera et al.\cite{54} have given the most important algorithm for all possible static locally anisotropic fluid solutions of the Einstein field equations in terms of two generating functions as:

\begin{equation}
 e^{\lambda(r)}=\frac{z^2\,e^{\int{\left[\frac{4}{r^2z(r)}+2\,z(r)\right]dr}}}
 {r^6\left[-2\,\int{\frac{z(r)(1+\Pi(r)\,r^2)e^{\int{\left[\frac{4}{r^2z(r)}+2\,z(r)\right]dr}}}{r^8}dr}+C\right]}
\end{equation}

where the two generating functions are: \\

$z(r)=\left[\frac{\nu^{\prime}}{2}+\frac{1}{r}\right]$, \, \, $\Pi=8\pi\,(p_r-p_t)$\\

According to the above algorithm, The generating functions in the present embedding class one case as follows (using the Eq.(10)):\\

\begin{equation}
z(r)=\frac{B\,\sqrt{e^{\lambda(r)}-1}}{A+B\,\int{\sqrt{e^{\lambda(r)}-1}\,dr}}+\frac{1}{r}
\end{equation}

\begin{equation}
\Pi=\frac{B\,\sqrt{e^{\lambda(r)}-1}}{2\,e^{\lambda}}\,\left[\frac{\,\sqrt{e^{\lambda(r)}-1}}{B\,r}-\frac{1}{A+B\,\int{\sqrt{e^{\lambda(r)}-1}\,dr}}\right]\,
\left(\frac{\lambda^{\prime}}{e^{\lambda}-1}-\frac{2}{r}\right).
\end{equation}

By using the Eq.(12), The Eqs. (21) and (22) yields the following generating functions $z(r)$ and $\Pi$ as:
\begin{equation}
z(r)=\frac{B\,b\,r\sqrt{a-b}}{\sqrt{1+br^2}\,[Ab+B\,\sqrt{a-b}\,\sqrt{1+br^2}]}+\frac{1}{r}
\end{equation}

\begin{equation}
\Pi={-ar^{2}\,\sqrt{(a-b)}}\,\left[\frac{A\,b\,\sqrt{(a-b)}\,\sqrt{1+br^{2}}+B\,(a-2b)\,(1+br^{2})}
{(1+ar^{2})^{2}\,\sqrt{1+br^{2}}\,[A\,b+B\,\sqrt{(a-b)}\,\sqrt{1+br^{2}}]}\right]
\end{equation}
Also in the present case the system is completely determined by providing one generating function through  $e^{\lambda(r)}$  and an additional ansatz in form of class one condition.

\section*{IV. PHYSICAL FEATURES OF THE COMPACT STAR:}

\noindent(a). Singularity at centre:\\

 (i). From Eqs. (12) and (13), we observe $e^{\lambda(0)}=1$ and $e^{\nu(0)}=\frac{1}{b^{2}}\,\left[A\,b+B\,\sqrt{a-b}\right]^{2}$. This shows that metric potentials are singularity free and positive at origin. Also it is monotonically increasing with increase the radius of the star (See Fig. 4).\\

 (ii).The energy density at centre $r=0$ implies that $\rho_{0}=\frac{3(a-b)}{8\pi}$.  Since density should be positive at
       Centre, then $a>b$.\\

\begin{figure}[!htp]\centering
	\includegraphics[width=5.5cm]{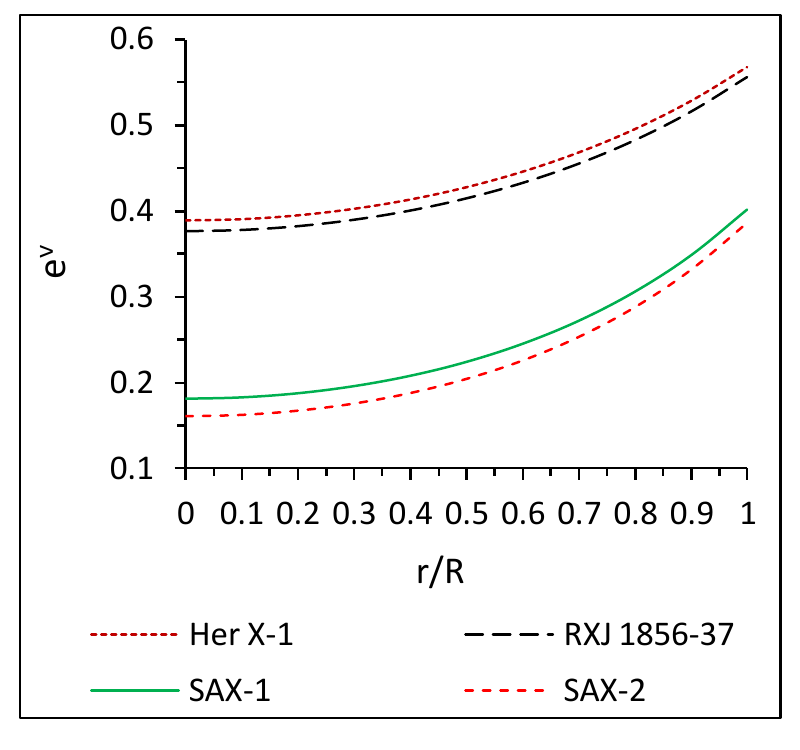}\includegraphics[width=5.5cm]{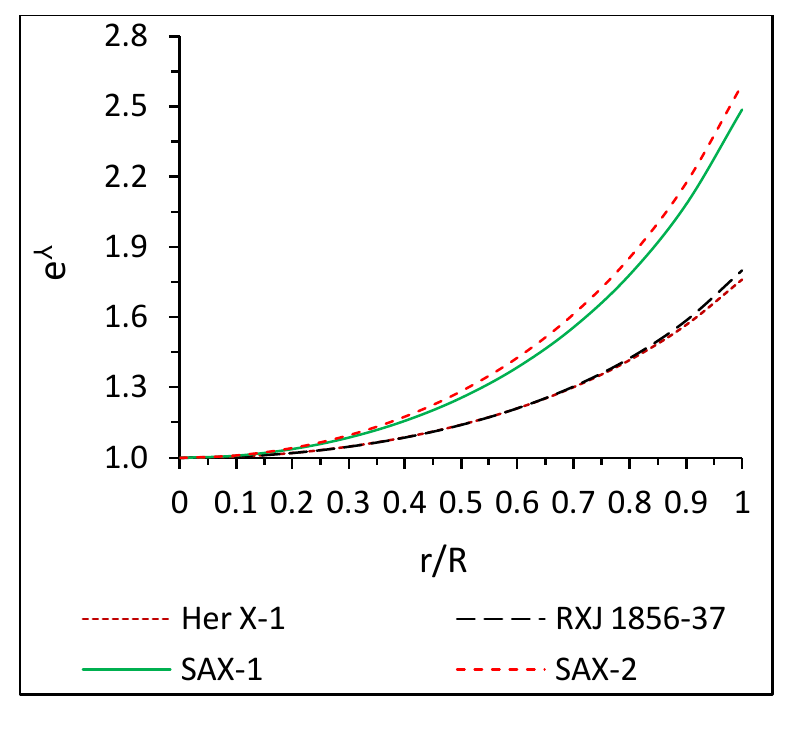}
		\caption{Variation of metric potentials $e^{\lambda }$ (left panel) and $e^{\nu }$ (right panel) with respect to fractional radius (r/R). For the purpose of plotting this figure, we have employed the same data set for the arbitrary constants: a, b, A and B with same mass and radius for each different compact star as in Figs. 1, 2 and 3}
		\label{Fig.4}
\end{figure}

\noindent (b). Velocity of Sound condition:\\

In the present models, the velocity of sound is monotonically decreasing away from centre and it is less than that velocity of light everywhere inside the compact star i.e. $0\le\,V_{i}=\sqrt{\frac{dp_{i}}{d\rho}}\le\,1$. According to the Canuto \cite{36}, the velocity of sound should decrease outwards for the EOS with an ultra-high distribution of matter. These features of velocity shows that our solution is well behaved and that behavior can be seen in Fig.(5).\\

\begin{figure}[!h]\centering
	\includegraphics[width=5.5cm]{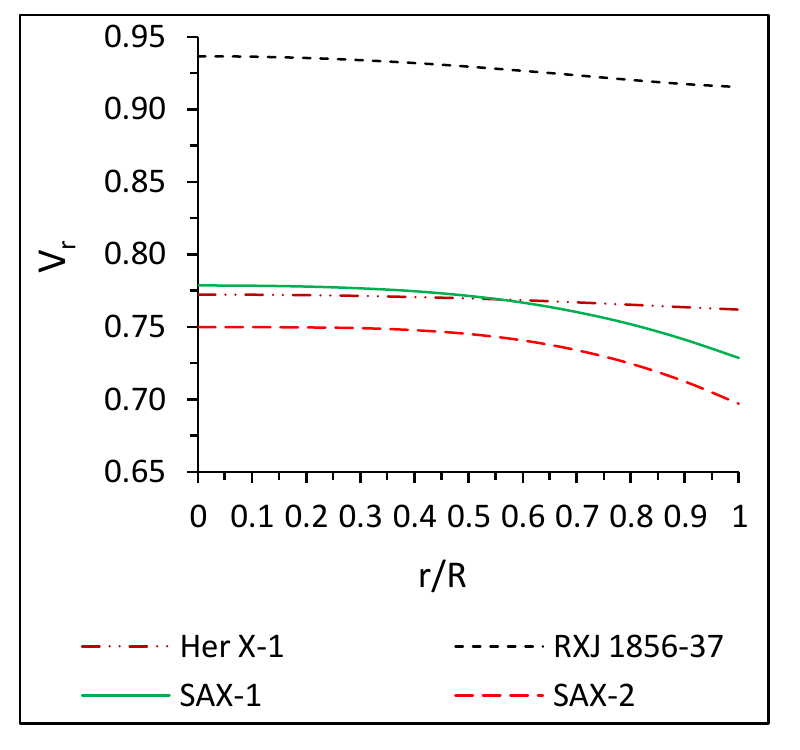}\includegraphics[width=5.5cm]{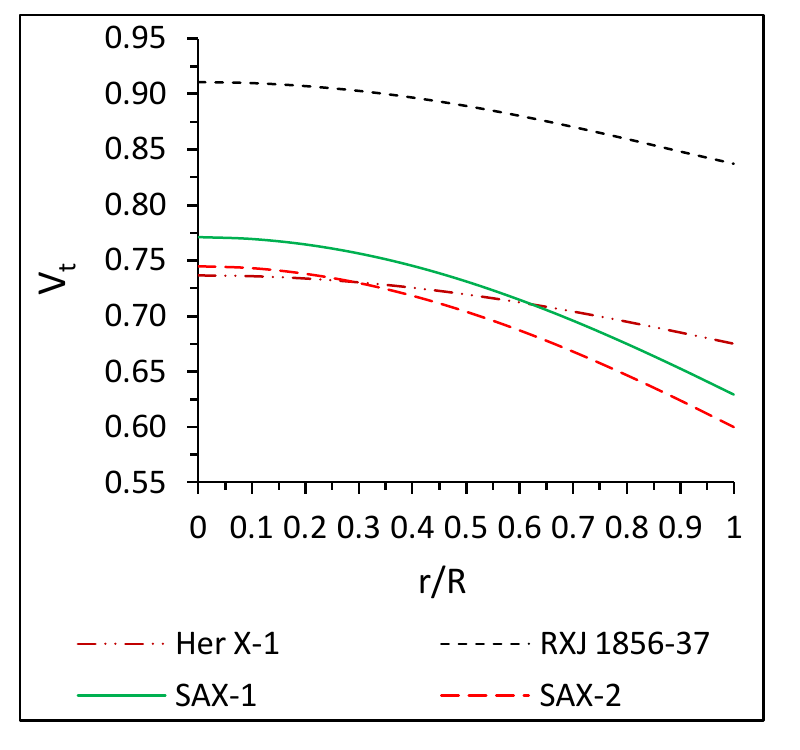}
	\caption{Variation of radial speed of sound (left panel) and transverse speed of sound (right panel) with respect to fractional radius (r/R).For this graph, the values of constants are as follows: (i). $a=0.0045, b=-0.0071, A=1.3704, B=0.0490$ with mass$(M)=0.9824 M_{\odot}$ and radius$(R)=6.7$ Km for $Her X-1$, (ii). $a=0.0041, b=-0.0101, A=1.2705, B= 0.0555$  and mass $(M)=0.9042 M_{\odot}$, radius$(R)=6.0$ Km for $RXJ 1856-37$, (iii). $a=0.0011, b=-0.0075, A=1.4134, B= 0.0547$, mass $(M)=1.4349 M_{\odot}$ and radius$(R)=7.07$ Km for $SAX J1808.4-3658(SS1)$ and (iv). $a=0.0017, b=-0.0087, A=1.5404, B= 0.0617$, Mass$(M)=1.3237 M_{\odot}$ and radius$(R)=6.37$ Km for $SAX J1808.4-3658(SS2)$}
	\label{Fig.5}
\end{figure}

\noindent (c). Behavior of functions $\omega_{r}$ and $\omega_{t}$: $\omega_{r}=\frac{p_{r}}{\rho}$,\, $\omega_{t}=\frac{p_{t}}{\rho}$ \\

The functions $\omega_{r}$ and $\omega_{t}$ is monotonically decreasing with the increase of $r$ i.e. $\left(\frac{d\omega_{t}}{dr}\right)_{r=0}$ , $\left(\frac{d\omega_{r}}{dr}\right)_{r=0}$  and $\frac{d^{2}\omega_{r}}{dr^{2}}$  is negative valued function for $r$ . This feature of $\omega_{r}$ and $\omega_{t}$  implies that the temperature of the compact star models decreases towards the surface of star. However both functions $\omega_{r}$ and $\omega_{t}$ is lying between 0 and 1. These behaviors are shown in Fig. 6.\\

\begin{figure}[!h]\centering
	\includegraphics[width=5.5cm]{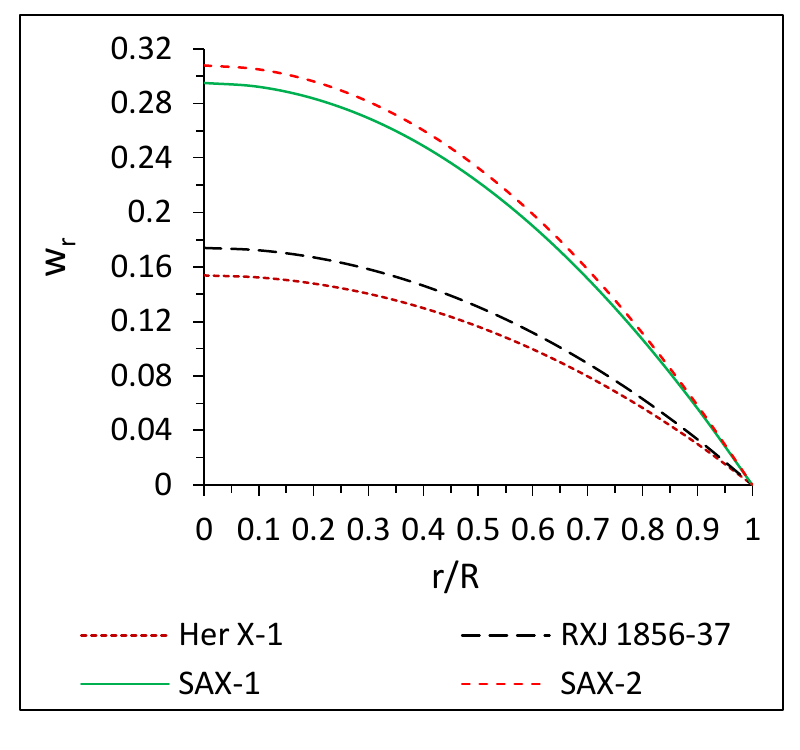}\includegraphics[width=5.5cm]{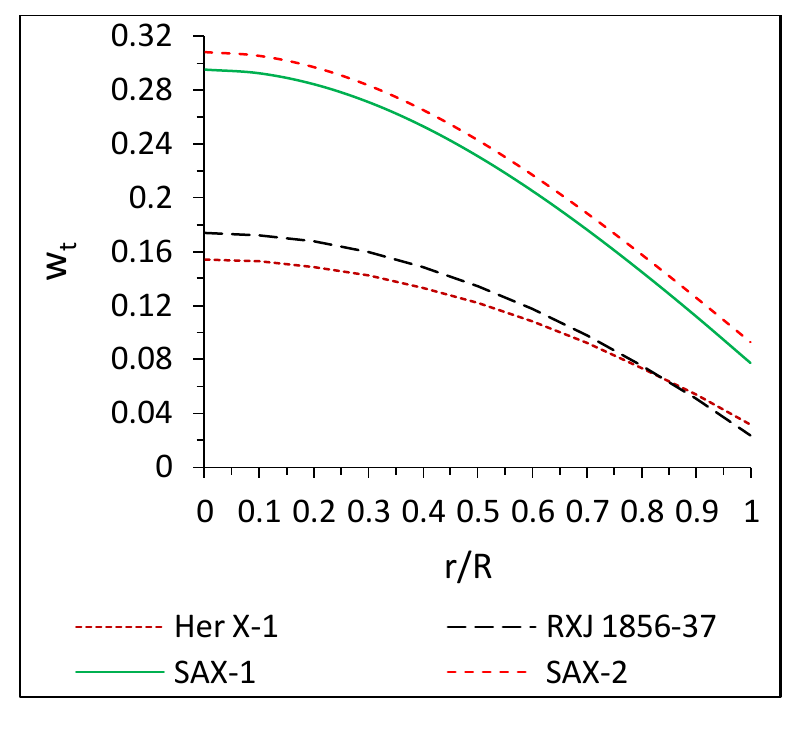}
	\caption{Variation of radial speed of sound (left panel) and transverse speed of sound (right panel) with respect to fractional radius (r/R).For the purpose of plotting this figure, we have employed the same data set for the arbitrary constants: a, b, A and B with same mass and radius for each different compact star as in Fig. 5}
	\label{Fig.6}
\end{figure}

\noindent (d). Matching Condition:\\

 To find out the expression for the constants A and B of the anisotropic model, we join smoothly the interior metric of anisotropic matter distribution to the exterior of Schwarzschild solution which is given by:
\begin{equation}
\label{20}
ds^{2} =\left(1-\frac{2M}{r} \right)\, dt^{2} -r^{2} (d\theta ^{2} +\sin ^{2} \theta \, d\phi ^{2} )-\left(1-\frac{2M}{r} \right)^{-1} dr^{2} ;
\end{equation}

Continuity of the metric coefficients $e^{\nu } $ and $e^{\lambda }$ across the boundary surface of the models  $r=R$ between the interior and the exterior regions of the star and radial pressure $ p_{r}=0$ at $r=R$ \cite{37}:\\

\noindent By Using above the boundary condition, we get:

\begin{equation}
\label{21}
 A=\frac{(3b-a)}{2b}\left[\frac{1+bR^{2}}{1+aR^{2}}\right]^{\frac{1}{2}};
\end{equation}

\begin{equation}
\label{22}
 B=\frac{\sqrt{a-b}}{2\,(1+aR^{2})^{\frac{1}{2}}};
\end{equation}

\begin{equation}
\label{23}
 M=\frac{R}{2}\,\left[1-\frac{1+bR^{2}}{1+aR^{2}}\right]
\end{equation}

\noindent (e). Energy conditions:\\

The models for anisotropic fluid distribution composed of strange matter should satisfy the following energy conditions: (i) Null energy condition (NEC), (ii) Weak energy condition (WEC) and (iii) Strong energy condition (SEC).
Then the following inequalities should hold simultaneously at all points inside the compact star corresponding to above conditions as:\\

\noindent NEC:  $\rho \ge 0$,\\
WEC${}_{r}$:     $\rho -p_{r} \ge 0$,\\
WEC${}_{t}$:      $\rho -p_{t} \ge 0$,\\
SEC:            $\rho -p_{r} -2p_{t} \ge 0$.\\

\begin{figure}[!htp]\centering
	\includegraphics[width=5.5cm]{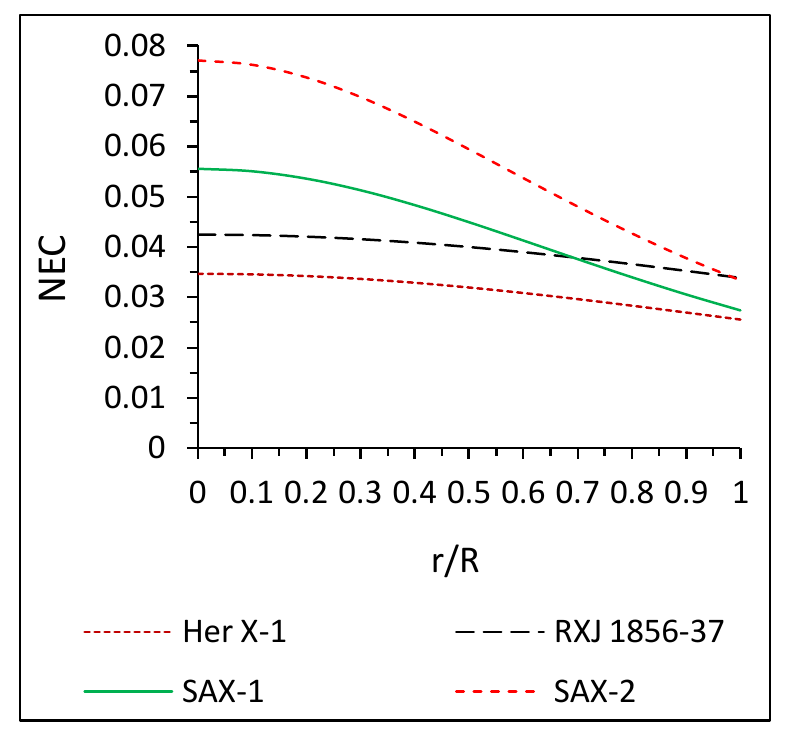}\includegraphics[width=5.5cm]{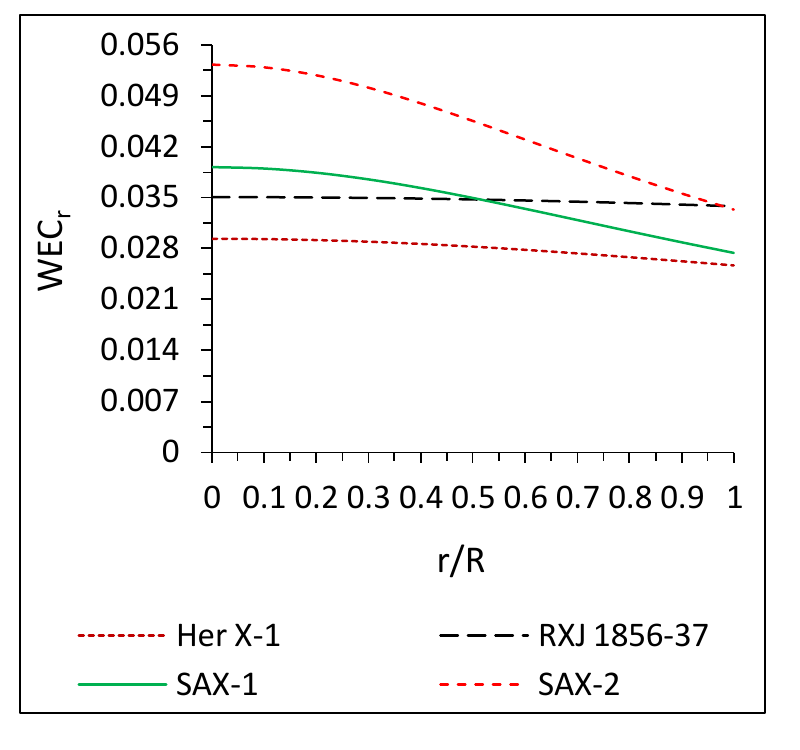}
     \includegraphics[width=5.5cm]{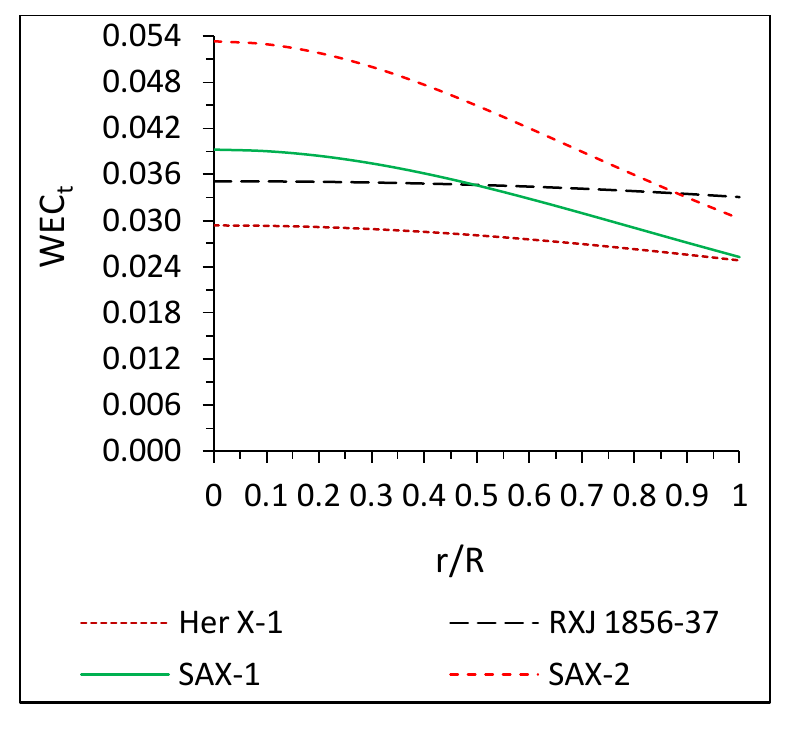}\includegraphics[width=5.5cm]{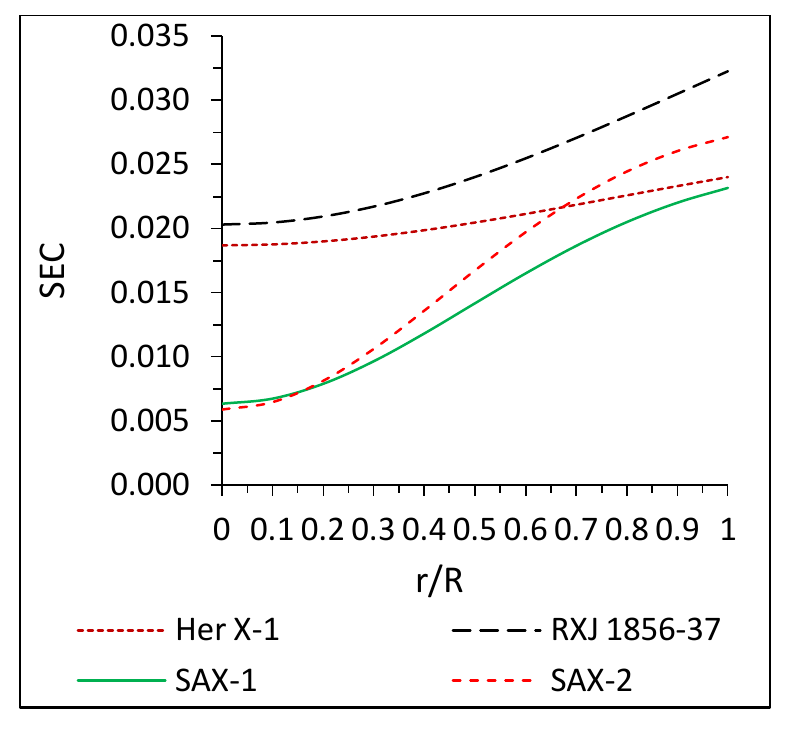}
	\caption{Variation of energy condition with radial coordinate r/R. (i). null energy condition (top left), (ii). radial weak energy condition (top right), (iii). transverse weak energy condition (bottom left),(iv). strong energy condition (bottom right). For the purpose of plotting this figure, we have employed the same data set for the arbitrary constants: a, b, A and B with same mass and radius for each different compact star as in Figs. 5 and 6}
	\label{Fig7}
\end{figure}

\noindent (f). Generalised Tolman-Oppenheimer -Volkoff (TOV) equation:\\

The generalized (TOV) equation for anisotropic fluid distribution is given by \cite{31,32},

\begin{equation}
\label{24}
\frac{M_{G} (\rho +p_{r} )}{r^{2} } e^{\frac{\lambda -\nu }{2} } +\frac{dp_{r} }{dr} +\frac{2}{r} (p_{r} -p_{t} )=0;
\end{equation}
\begin{equation}
\label{25}
-\frac{1}{2}\,\nu'(\rho +p_{r} )-\frac{dp_{r} }{dr} +\frac{2}{r} (p_{t} -p_{r} )=0;
\end{equation}

Where $M_{G}$ is the effective gravitational mass given by
\begin{equation}
\label{26}
M_{G} (r)=\frac{1}{2} r^{2} e^{\frac{\nu -\lambda }{2} } \, \nu '
\end{equation}
The Eq.(30) describes the equilibrium condition for an anisotropic fluid distribution subject to gravitational ($F_{g}$), hydrostatic ($F_{h}$) and anisotropic stress ($F_{a}$) so that:
\begin{equation}
\label{27}
F_{g} +F_{h} +F_{a} =0 ,
\end{equation}
where, its components can be defined as:
\begin{equation}
\label{28}
F_{g} =-\frac{1}{2} \nu '\, (\rho +p_{r} )
\end{equation}
\begin{equation}
\label{29}
F_{h} =-\frac{dp_{r} }{dr}
\end{equation}
\begin{equation}
 \label{30}
F_{a} =\frac{2}{r} (p_{t} -p_{r} )
\end{equation}

The components for forces can be expressed in explicit form as:
\begin{equation}
 \label{31}
F_{g} =\frac{(a-b)\,B\,b\,r}{4\pi}\,\left[\,\frac{A\,b\,\sqrt{(a-b)}+a\,B\,(1+br^{2})^{\frac{3}{2}}}{(1+ar^{2})^{2}\,\sqrt{1+br^{2}}\,[\,A\,b+B\,\sqrt{(a-b)}\,\sqrt{1+br^{2}}\,]^{2}}\right]
\end{equation}

\begin{equation}
 \label{32}
F_{h} =-\frac{r\,\sqrt{(a-b)}}{4\pi}\left[\,\frac{a\,A^{2}b^{2}\sqrt{(a-b)}+a\,B^{2}\sqrt{(a-b)}\,(a-3b)\,(1+br^{2})+F_{1}(r)}{(1+ar^{2})^{2}\,[\,A\,b+B\,\sqrt{(a-b)}\,\sqrt{1+br^{2}}\,]^{2}}\right]
\end{equation}

\begin{equation}
 \label{33}
F_{a} =\frac{a\,r\sqrt{(a-b)}}{4\pi}\,\left[\,\frac{A\,b\,\sqrt{(a-b)}\,\sqrt{1+br^{2}}+\,B\,(1+br^{2})\,(a-2b)}{(1+ar^{2})^{2}\,\sqrt{1+br^{2}}\,[\,A\,b+B\,\sqrt{(a-b)}\,\sqrt{1+br^{2}}\,]}\right]
\end{equation}\\

where,\, \, $F_{1}(r)=a\,A\,B\,[\,b^{2}-ab\,(4+3br^{2})+2a^{2}(1+br^{2})]$.\\

Fig. 8 represents the behavior of generalized TOV equations. We observe from this figure that the system is counter balance by the components the gravitational force ($F_{g}$), hydrostatic force ($F_{h}$) and anisotropic stress ($F_{a}$) and the system attains a static equilibrium.
\begin{figure}[!htp]\centering
	\includegraphics[width=5.5cm]{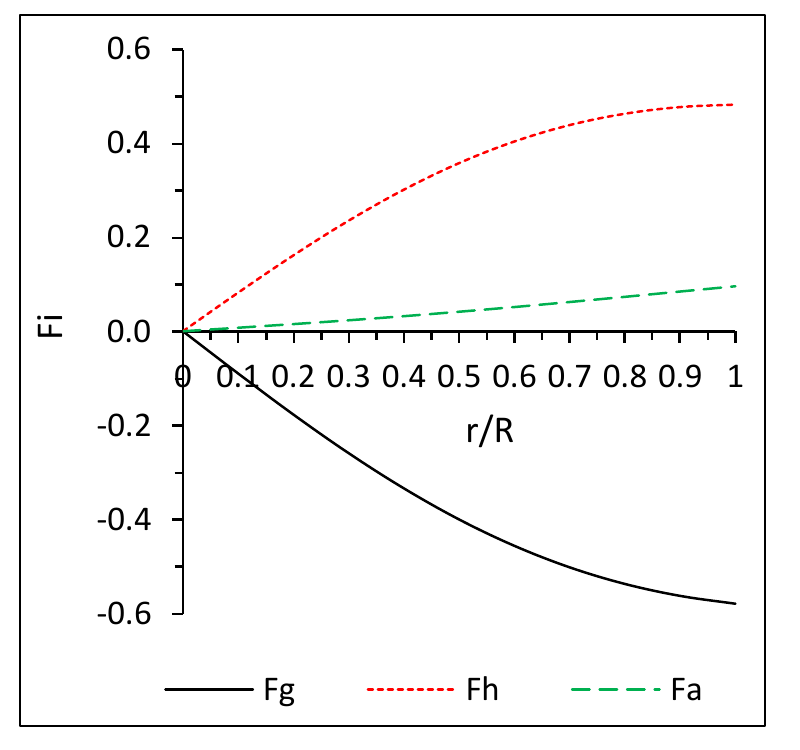}\includegraphics[width=5.5cm]{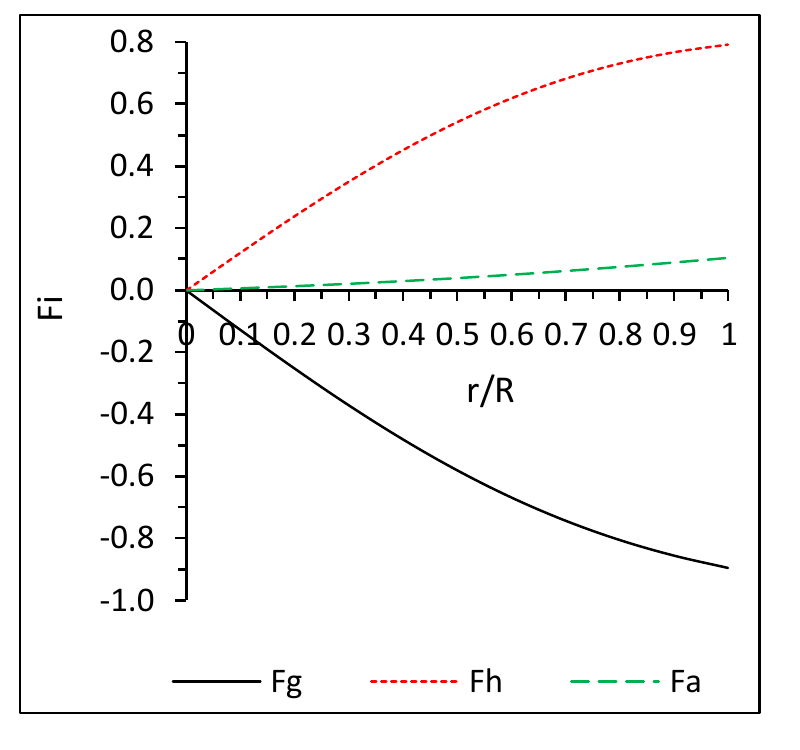}
     \includegraphics[width=5.5cm]{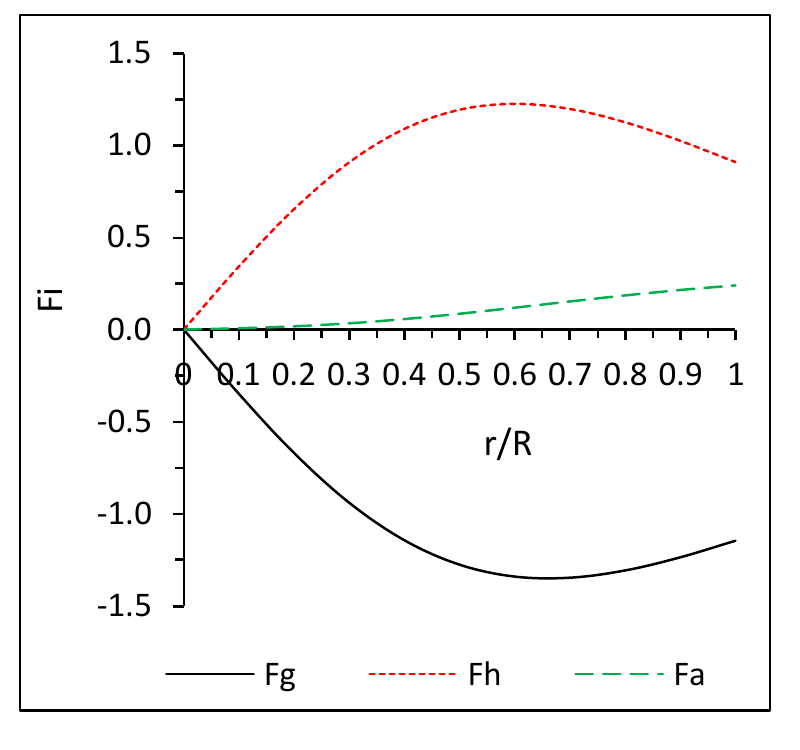}\includegraphics[width=5.5cm]{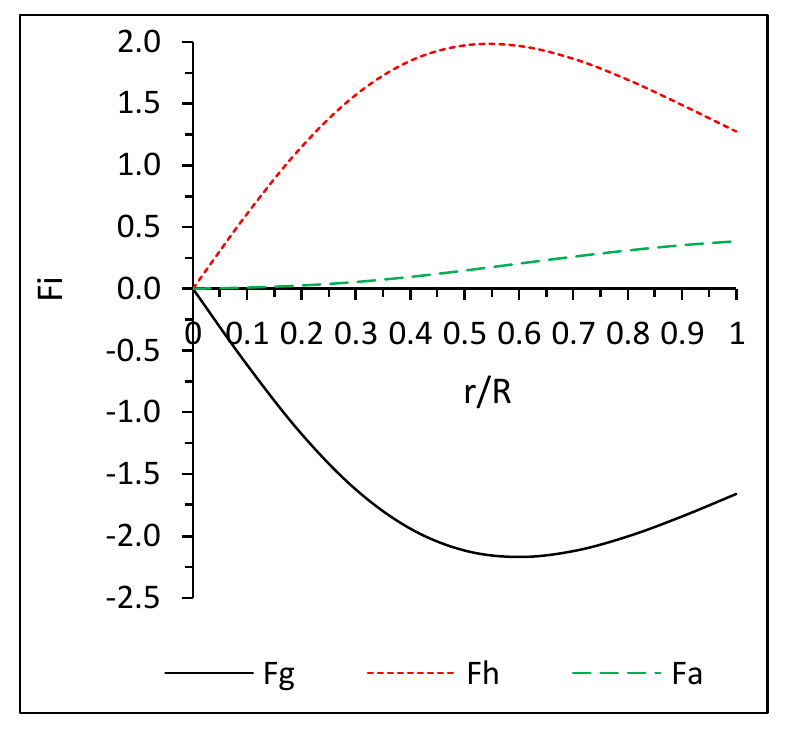}
	\caption{Variation of different forces with radial coordinate  r/R. (i). Her X-1 (Top left) with value of constants: $a=0.0045, b=-0.0071, A=1.3704, B=0.0490$ and mass $(M)=0.9824 M_{\odot}$, radius$(R)=6.7$ Km. (ii). RXJ 1856 - 37 (Top right) with value of constants:  $a=0.0041, b=-0.0101, A=1.2705, B= 0.0555$ and mass $(M)=0.9042 M_{\odot}$, radius $(R)=6.0$ Km. (iii). SAX J1808.4 - 365(SS1) (bottom left)with values of constants: $a=0.0011, b=-0.0075, A=1.4134, B= 0.0547$ and mass $(M)=1.4349 M_{\odot}$, radius$(R)=7.07$  (iv). SAX J1808.4 - 3658 (SS2) (Bottom left) with value of constants: $a=0.0017, b=-0.0087, A=1.5404, B= 0.0617$ and mass$(M)=1.3237 M_{\odot}$ , radius $(R)=6.37$.}
	\label{Fig.8}
\end{figure}

However, the gravitational force is dominating the hydrostatic force and it is balanced by the joint action of hydrostatic force and anisotropic stress while the anisotropic stress has a less role to the action of equilibrium condition. These physical features represents that the models are stable.

\section*{V. STABILITY  OF THE COMPACT STAR MODELS:}
\noindent For physically acceptable anisotropic fluid stellar models, the velocity of sound must be less than that velocity of lights i.e. it should be within the range $0\le\,V_{i}=\sqrt{\frac{dp_{i}}{d\rho}}\le\,1$. Since Causality condition for radial and transverse velocity of sound is less than 1 this implies that $0\le\,V_{i}^{2}=\frac{dp_{i}}{d\rho}\le\,1$. The plot for square velocity of sound is shown in Fig.9, we observe that it is monotonically decreasing and less than 1 throughout inside the star.
Now using the cracking concept of Herrera and Abreu et al. \cite{29,30}, to determine the stability of local anisotropic fluid models, which states that the region is potentially stable where the radial velocity of sound is greater than the transverse velocity of sound.
\begin{figure}[!h]\centering
	\includegraphics[width=5.5cm]{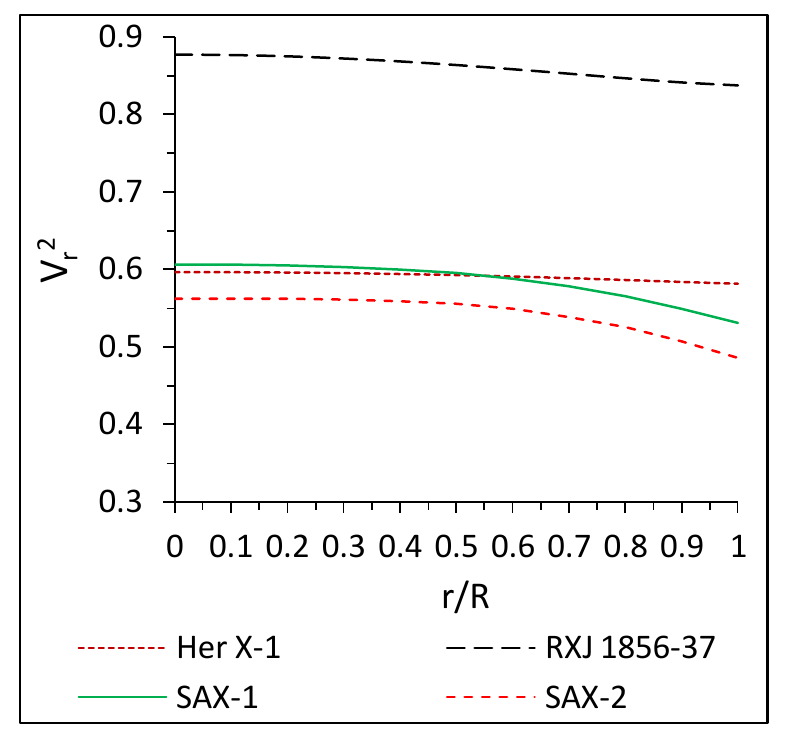}\includegraphics[width=5.5cm]{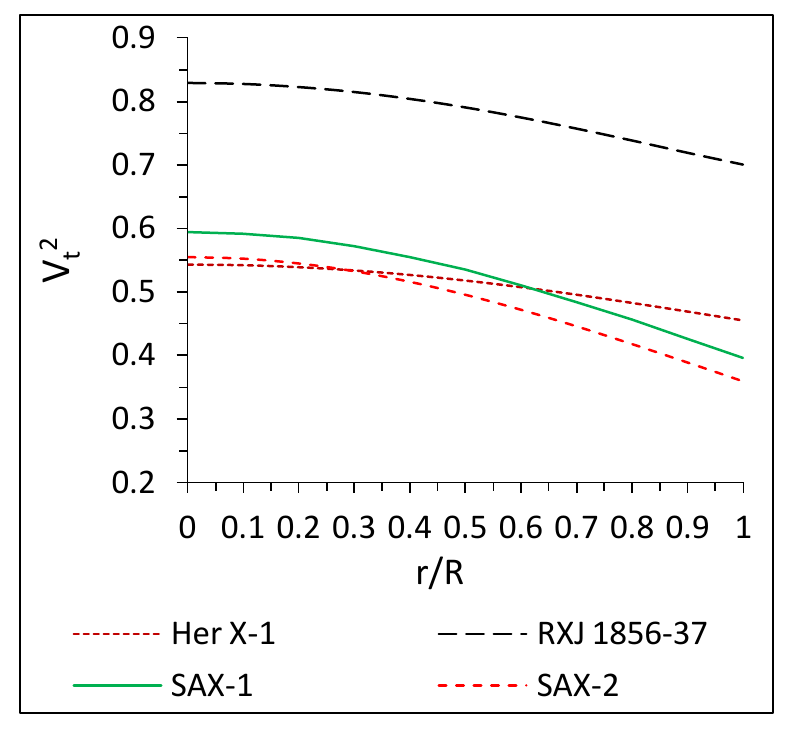}
	\caption{Variation of square of radial speed of sound (left panel)and transverse speed of sound (right panel) with respect to fractional radius (r/R). For the purpose of plotting this figure, we have employed the same data set for the arbitrary constants: a, b, A and B with same mass and radius for each different compact star as in Fig. 8}
	\label{Fig.9}
\end{figure}
Then the absolute difference between radial and transverse speed of sound is given as:
\begin{equation}
 \label{34}
V_{t}^{2}-V_{r}^{2}=\sqrt{(a-b)}\,\left[\frac{2\,\sqrt{1+br^{2}}\,\sqrt{(a-b)}\,(ar^{2}-1)\,[\,A^{2}\,b^{2}+B^{2}(1+br^{2})\,(a-2b)\,]-f(r)}{2\,(a-b)(5+ar^{2})^{2}\,\sqrt{1+br^{2}}\,[\,A\,b+B\,\sqrt{(a-b)}\,\sqrt{1+br^{2}}]^2}\right]
\end{equation}\\

where, \, \, $f(r)=b\,A\,B\,[(4a-6b)+br^{2}(10a-7b)-ar^{2}(4a+4abr^{2}-5b^{2}r^{2})]$.\\

\begin{figure}[!h]\centering
	\includegraphics[width=5.5cm]{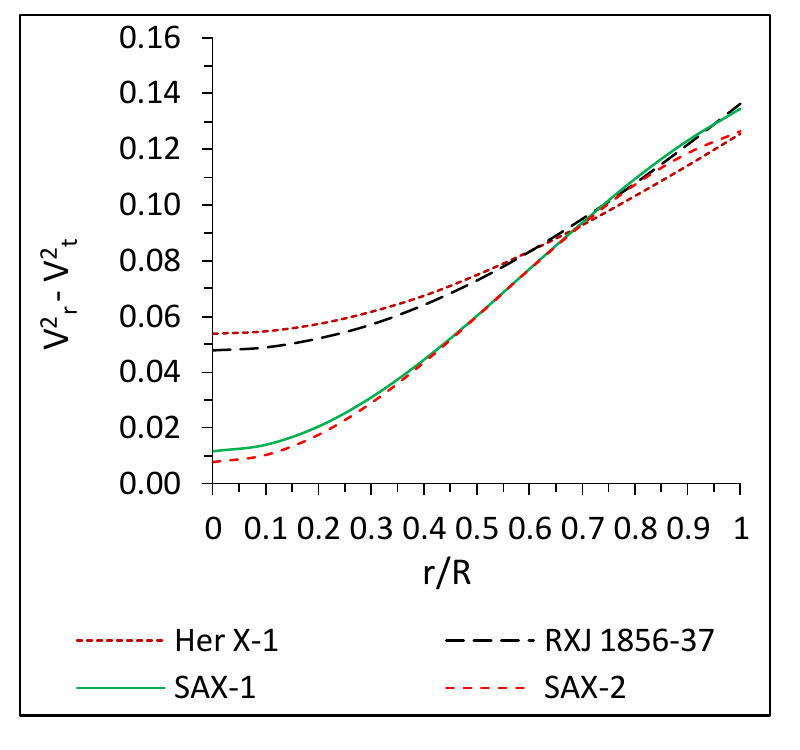} \includegraphics[width=5.5cm]{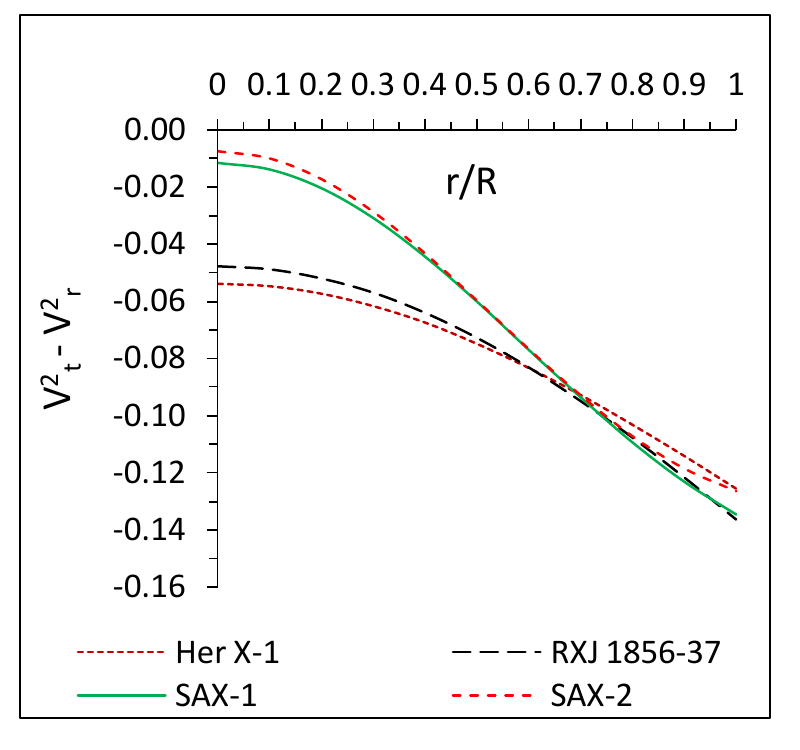}
	\caption{Variation of the  $V_{r}^{2}-V_{t}^{2}$  and  $V_{t}^{2}-V_{r}^{2}$  value of square of sound velocity with respect to fractional radius (r/R). For the purpose of plotting this figure, we have employed the same data set for the arbitrary constants: a, b, A and B with same mass and radius for each different compact star as in Figs. 8 and 9}
	\label{Fig.10}
\end{figure}
The Fig.10 indicates that  radial velocity of sound is greater than the transverse velocity of sound which immediately confirms that our model of anisotropic compact stars are stable.

\section*{VI. RELATION BETWEEN EFFECTIVE MASS WITH RADIUS AND SURFACE RED SHIFT:}

\noindent In present section, we will discuss the maximum allowable mass in our stellar models. According to Buchdahl \cite{38}, the maximum limit of mass-radius ratio for static spherically symmetric perfect fluid star is. For more generalized expression for the same mass-radius ratio can be seen in \cite{22}.

\noindent  For compactness, the expression for ratio of effective mass and radius is define as:

\begin{equation}
 \label{35}
u=\frac{M_{eff}}{r}=\frac{1}{2}\,\left[\frac{(a-b)R^{2}}{1+aR^{2}}\right]
\end{equation}
\begin{figure}[!h]\centering
	\includegraphics[width=5.5cm]{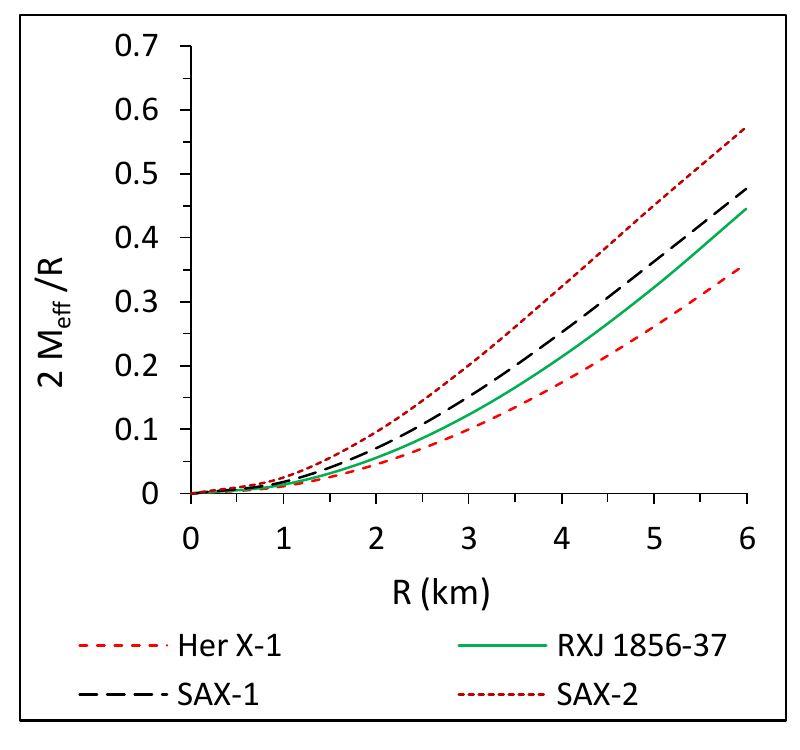}
	\caption{Variation of the $\frac{2M_{eff}}{R}$  with respect to radius (R).}
	\label{Fig.11}
\end{figure}
From the Fig. (11), it is clear that the ratio $\frac{2M_{eff}}{R}$ is monotonically increasing with the radius and satisfying the Buchdahal's limit (Table 2).

\begin{table}
	\centering
	\caption{Values of the model parameters Mass ($M_{\odot}$), radius ($R$), mass-radius ratio ($M/R$) and arbitrary constants $a$,\,$b$,\,$A$ and $B$ for different compact stars:}
	\label{Table1}
\begin{tabular}{@{}lrrrrrrr@{}} \hline
Compact star\\ candidates & $M(M_{\odot})$ & $R$(Km) & $M/R$ & $a(km^{-2}) $ & $b(km^{-2})$ & $A$ & $B(km^{-1})$ \\ \hline
Her. X-1                              & 0.9824 & 6.7  & 0.216  & 0.0045 & -0.0071 & 1.3704 & 0.0490 \\ \hline
RXJ 1856-37                           & 0.9042 & 6.0  & 0.222  & 0.0041 & -0.0101 & 1.2705 &  0.0555 \\ \hline
SAX J1808.4-3658(SS1)                 & 1.4349 & 7.07 & 0.299  & 0.011  & -0.0075 & 1.4134 & 0.0547 \\ \hline
SAX J1808.4-3658(SS2)                 & 1.3237 & 6.35 & 0.3071 & 0.017  & -0.0087 & 1.5404 &  0.0617   \\ \hline
\end{tabular}
\end{table}

The surface red-shift ($Z_{s}$) corresponding to the above compactness ($u$) is obtained as:

\begin{equation}
 \label{36}
Z_{s}=\frac{1-[1-2u]^{\frac{1}{2}}}{[1-2u]^{\frac{1}{2}}}=\frac{(a-b)R^{2}}{[\,1+bR^{2}+\sqrt{1+(a+b)R^{2}+abR^{4}}\,]}.
\end{equation}

\begin{figure}[!h]\centering
	\includegraphics[width=5.5cm]{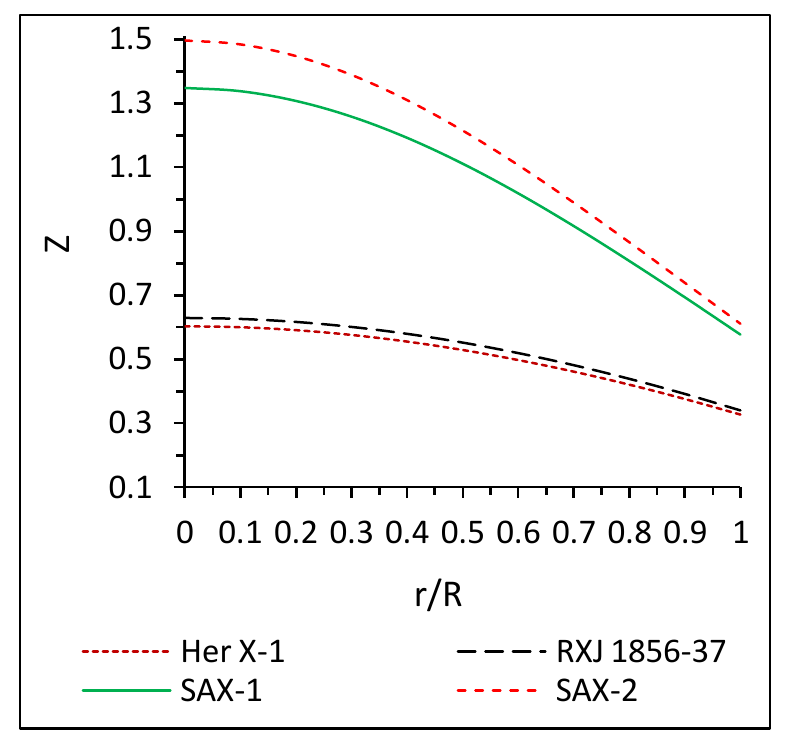}
	\caption{Variation of red shift ($Z$) with respect to fractional radius (r/R). we have employed the same data set for the arbitrary constants: a, b, A and B with same mass and radius for each different compact star as in Figs. 8, 9 and 10}
	\label{Fig.12}
\end{figure}

\begin{table}
	\centering
	\caption{Energy densities, central pressure and Buchdahl condition for different compact star candidates for the above parameter values of Tables 1 }
	\label{Table2}
\begin{tabular}{@{}lrrrr@{}} \hline
Compact star & Central Density & Surface density & Central pressure & Buchdahl  \\
candidates & $gm/cm^{3} $ & $gm/cm^{3}$ & $dyne/cm^{2} $ & condition\\\hline
Her. X-1  & 1.8635$\times 10^{15} $ & 1.3766$\times 10^{15} $ & 2.5819$\times 10^{35} $ &$2M/R=0.432< 8/9$  \\ \hline
RXJ 1856-37 & 2.2802$\times 10^{15} $ & 1.8165$\times 10^{15} $ & 3.5695$\times 10^{35}$ & $2M/R=0.444<8/9$  \\ \hline
SAX J1808.4-3658(SS1) & 2.9871$\times 10^{15} $ & 1.4715$\times 10^{15} $ & 7.9374$\times 10^{35}$ & $2M/R=0.598< 8/9$  \\ \hline
SAX J1808.4-3658(SS2) & 4.1360$\times 10^{15} $ & 1.7886$\times 10^{15} $ & 11.4630$\times 10^{35}$ & $2M/R=0.6142< 8/9$  \\ \hline
\end{tabular}
\end{table}
\section*{VII. COMPARISION BETWEEN OBSERVATIONAL OBJECTS WITH THE PRESENT MODELS:}
 \noindent We now point out about the   estimate of the range of various physical parameters of different strange star candidates like Her X-1, RXJ 1856-37, SAX J 1808.4- 3658 (SS1), SAX J 1808.4- 3658(SS2).  We have calculated the values of the relevant physical parameters by assuming the estimated mass and radius of these stars   which have been given in Table 1.  Plugging in $G$ and $c$ in the relevant equations, we have calculated central density and pressure as well as surface density. The estimated numerical values for different cases have been shown in Table 2. Also we have evaluated the numerical values of red shift for different compact star candidates which have been shown in Table 3. The estimated physical parameters of different characteristics of the compact stars what we have shown here is mathematically consistent and physically reasonable analytic solution for modeling compact stars.

 \begin{table}
	\centering
	\caption{Numerical Values of red-shift ($Z$) for different compact star models}
	\label{Table3}
	\begin{tabular}{cccccc|ccccc} \hline
		$r/R$ &Her X-$1$ &RXJ $1856-37$ &SAX-$1$ &SAX-$2$  \\ \hline
		0.0 & 0.6026 & 0.6301 & 1.3474 & 1.4974 \\ \hline
		0.1 & 0.5996 & 0.6270 & 1.3372 & 1.485  \\ \hline
		0.2 & 0.5905 & 0.6176 & 1.3069 & 1.4485 \\ \hline
		0.3 & 0.5755 & 0.6020 & 1.2578 & 1.3895   \\ \hline
		0.4 & 0.5548 & 0.5804 & 1.1918 & 1.3107   \\ \hline
		0.5 & 0.5286 & 0.5531 & 1.1111 & 1.2154  \\ \hline
		0.6 & 0.4972 & 0.5203 & 1.0184 & 1.1071  \\ \hline
		0.7 & 0.4610 & 0.4825 & 0.9162 & 0.9893  \\ \hline
		0.8 & 0.4203 & 0.4398 & 0.8072 & 0.8654  \\ \hline
		0.9 & 0.3755 & 0.3926 & 0.6936 & 0.7382  \\ \hline
		1.0 & 0.3269 & 0.3411 & 0.5772 & 0.6100  \\ \hline
	\end{tabular}
\end{table}
\section*{VIII. DISCUSSIONS AND CONCLUSIONS:}
   \noindent We present a very simple and new unique anisotropic compact star models of Einstein field equation for spherically symmetric metric in class one. The space-time same as Maurya et al.\cite{39}in which we have determined a electromagnetic mass model which describe a model of electron. However in the present paper we have utilized the this space time for anisotropic fluid distribution without electric charge. Here we have shown through the solution set that the central energy density $\rho_{0}$ is non zero and free from singularity at the centre $r=0$. Also both central and surface density has the order of $10^{15}gm/cm^{3}$. This implies that our solution can represent the realistic astrophysical compact star models. However the physical parameters of the stars in dependent on anisotropy factor i.e. the energy density and pressures of the compact star are present in the interior of the star due to pressure anisotropy. We can say the interior of the present stars are purely anisotropic.\\
The physical features for the interior of the star can be explained as follows:\\
\noindent (i).The energy density and pressures of the stars are positive and finite everywhere inside the stars (Figs. 1 and 2).\\
\noindent (ii).The anisotropic star models are satisfying the various energy conditions as mentioned in Sec.IV(e) (Fig.7).\\
\noindent (iii). We observe from Fig.10, the radial speed of sound is always greater than transverse speed of sound everywhere inside the stars because there is no change in sign of $v_{r}^{2}-v_{t}^{2}$ and $v_{t}^{2}-v_{r}^{2}$.Therefore our compact star models are stable [29,30].\\
\noindent (iv). We have calculated the effective mass of each present obtained compact star like Her. X-1, RXJ $1856-37$, SAX J$1808.4-3658$(SS1) and SAX J$1808.4-3658$(SS2) and observe that it is monotonically increasing with the radius of the star (Fig.11) and satisfy the Buchdahal upper limit.\\
\noindent (v).The red-shift is monotonically decreasing. However it is maximum at centre and minimum at the boundary of the star (Fig.12 and Table 3). For isotropic case without cosmological constant, the red-shift $Z\le 2$   \cite{30,40,41}. Bohmer and Harko \cite{41} argued that for an anisotropic star with cosmological constant the surface red shift must satisfy general restriction $Z\le 5$. From Table(3), it is clear that the redshift of the present compact star models are in good agreement.\\

\textbf{Acknowledgments}: The authors S. K. Maurya et al. acknowledge continuous support and encouragement from the administration of University of  Nizwa. Also We all are thankful to the anonymous referee for raising several pertinent issues, which have helped us to improve the manuscript substantially.

\end{document}